# Hyperhoneycomb boron nitride with anisotropic mechanical, electronic and optical properties

Jin Yu[1,2], Lihua Qu[3], Edo van Veen[2], Mikhail I. Katsnelson[2] and Shengjun Yuan[4,1,2*]

1 Beijing Computational Science Research Center, Beijing 100094, China

2 Theory of Condensed Matter, Radboud University, Heyendaalseweg 135, 6525 AJ Nijmegen, The Netherlands

3 School of Science, Nantong University, Nantong 226019, China

4 School of Physics and Technology, Wuhan University, Wuhan 430072, China

## Abstract

Boron nitride structures have excellent thermal and chemical stabilities. Based on state-of-art theoretical calculations, we propose a wide gap semiconducting BN crystal with a three-dimensional hyperhoneycomb structure (Hp-BN), which is both mechanically and thermodynamically stable. Our calculated results show that Hp-BN has a higher bulk modulus and a smaller energy gap as compared to c-BN. Moreover, due to the unique bonding structure, Hp-BN exhibits anisotropic electronic and optical properties. It has great adsorption in the ultraviolet region, but it is highly transparent in the visible and infrared region, suggesting that the Hp-BN crystal could have potential applications in electronic and optical devices.

**TOC**

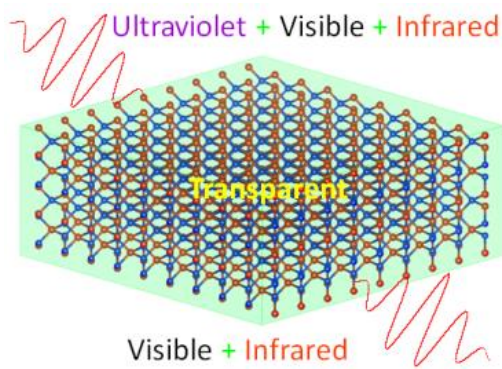

**Introduction**

Carbon-based materials have attracted continuous interest in the last decade because of the rise of graphene.[1] As the left and right neighbors of carbon in the elementary table, boron and nitrogen can be isomorphic to carbon in any lattice structure, such as a zero-dimensional (0D) cage,[2-4] one-dimensional (1D) nanotube,[5] two-dimensional (2D) monolayer,[6, 7] and three-dimensional (3D) diamond-like crystal structure.[8-10] Unlike purely covalent C-C bonds in carbon-based materials, the chemical bonds in BN binary materials show distinct ionicity because of the localized electronic states induced by different electronegativity between boron and nitrogen atoms, which are superior chemically stable.[11] Thus, BN structures are widely used as high-temperature ceramic materials in extremely harsh environments. On the other hand, the ionicity breaks the symmetry of the electronic states and opens a large band gap in BN materials, which makes them insulating. All currently identified 3D BN phases in experiment are insulators, including cubic-BN (c-BN),[9] wurtzite-BN (w-BN)[10] and layered hexagonal BN (h-BN).[7] Recently, theoretical investigations have shown that the insulating properties of BN materials can be modulated by introducing a hybrid $sp^2$ bond. A thermodynamically stable phase of BN (T-$B_xN_x$), partially hybrid by $sp^2$ and $sp^3$ bonds, has been reported to be metallic due to delocalized boron $2p$ electrons.[12] 3D BN networks containing only $sp^2$ hybrid BN bonds are reported to be semiconducting as well.[13] However, semiconducting 3D BN with purely $sp^3$ hybrid bonds has never been reported before.

In this work, we report a new allotrope of BN with a unique 3D hyperhoneycomb structure, using an unbiased particle swarm optimization (PSO) structure search algorithm.[14] Based on comprehensive density functional calculations, we found that the structure of the proposed BN (Hp-BN) is thermodynamically and elastically stable. The calculated bulk modulus of Hp-BN lies between conventional c-BN and w-BN. Moreover, Hp-BN is semiconducting with an indirect energy gap around 3.45 eV. Due to the unique hybridization of $p$ orbitals in B and N atoms, it exhibits anisotropic optical properties with birefringence index. It has great adsorption in the ultraviolet region and is highly transparent in the visible and infrared region, which gives Hp-BN potential applications in optical devices.

**Computational methods**

The candidate structure was obtained using a global structural optimization (GSO) method, as implemented in CALYPSO code,[14-16] which has been successfully applied in material investigations.[17-20] The subsequent structural relaxation and total energy calculations were carried out using density functional theory (DFT) as implemented in the Vienna ab initio simulation package (VASP).[21, 22] Electron exchange and correlation interactions were described using the Perdew-Burke-Ernzerhof (PBE) pseudopotentials within the general gradient approximation (GGA).[23] The plane wave cut off energy was set to 500 eV. The Brillouin Zone sampling[24] was done using a 15*15*13 Monkhorst-Pack grid for relaxation calculations and a 51 *k*-point sampling of Line-Mode was used for the static calculations. All the atoms in the unit cell were fully relaxed until the force on each atom was less than 0.01 eV/A. Electronic minimization was performed with a tolerance of $10^{-5}$ eV. To ensure an accurate determination of electronic properties, calculations were repeated using the hybrid Heyd-Scuseria-Ernzerhof functionl (HSE06).[25] The many-body quasiparticle energies were calculated in the self-consistent GW approximation ($G_0W_0$) as implemented in VASP.[26] The photon energy dependent dielectric function was obtained by solving the Bethe-Salpeter equation for the two-particle Green's function within the Tamm-Dancoff approximation as implemented in VASP.[27-29] Mulliken population analysis was performed by using CASTEP code after full structure optimization.[30] Lattice vibrational properties were calculated using the density functional perturbation theory (DFPT) with a 3*3*2 supercell.[31] Before calculating the phonons, the structures were reoptimized using a higher convergence criteria of $10^{-8}$ eV for total energy and $10^{-6}$ eV/A for Hellmann-Feynman Force, respectively. Phonon band and frequency DOS

were obtained by solving the dynamical equations using the PHONOPY code.[32]

**Results**

The calculated cohesive energies of 240 BN structures from CALYPSO are shown in Figure 1. In the structure search, the BN formula unit varies from 1:1 to 4:4 with a fixed stoichiometric composition of 1:1. The cohesive energies of most BN nanostructures are located between -9 eV/atom and -5 eV/atom. The density of states (DOS) on the right panel shows that a sharp peak appears around -8.75 eV/atom, which corresponds to a frequency that appears in most stable BN structures, including layered h-BN sheets (-8.79 eV/atom) and the well-known c-BN (-8.71 eV/atom). Both are in good agreement with previous studies. The cohesive energy of the proposed Hp-BN structure lies in the second sharpest peak of the DOS with the value being -7.78 eV/atom, indicating that it is metastable as compared to c-BN. However, it is much lower than many BN structures reported previously, such as rocksalt,[33] NiAs-type,[34] cage-like structures[2, 3] and T-$B_xN_x$,[12] as well as some BN nanotubes,[12] which shows that Hp-BN is much more stable.

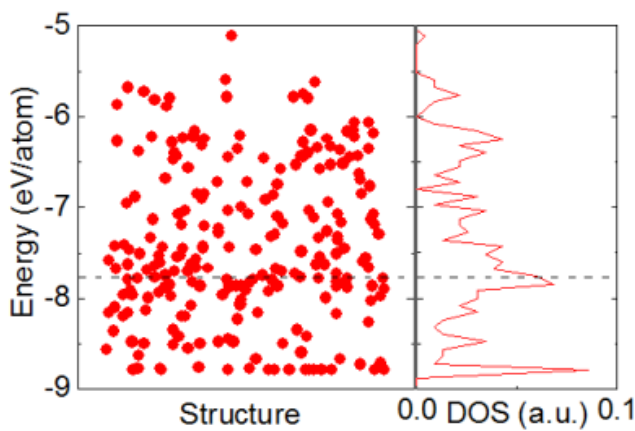

**Figure 1** Schematic of the calculated cohesive energies of various BN structures. The red dots represent different BN structures obtained from CALYPSO. The dashed line indicates the cohesive energy of the proposed Hp-BN.

Figure 2 shows that the atomic configuration of the Hp-BN structure is designed with an orthorhombic primitive cell (space group P6$_2$22, NO. 180) containing three formula units. The optimized lattice parameters are $a = b$ = 2.610 Å and $c$ = 5.828 Å. To simplify the geometric structure, we label the B and N atoms in Hp-BN from the top to the bottom of the unit cell as B1, N1, B2, N2, B3 and N3, respectively. Their atomic Wyckoff positions are given in Table S1 in the Supporting Information. Each B (N) atom in the hyperhoneycomb binds to four N (B) atoms with a bond length of 1.627 Å, which gives a reasonable explanation for the nearly 0.9 eV/atom difference of the cohesive energy between Hp-BN and c-BN (with a bond length of 1.569 Å). The hyperhoneycomb structure can be further divided into three vertical planes by the B1-N1, B2-N2 and B3-N3 bonds, respectively. In each plane, the structure can be viewed as zigzag BN nanoribbon (ZBNNR) structures, with a 106.68° bond angle between B-N-B (N-B-N). Along the $c$ direction, those ZBNNR-like structures are connected by the N1-B2, N2-B3 and N3-B1 bond with a 60° in-plane rotation angle.

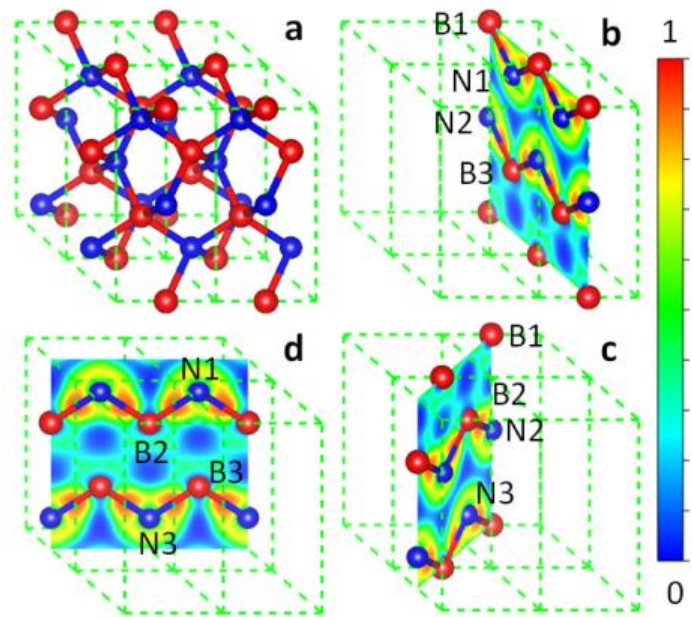

**Figure 2** Atomic configuration and slices of the electron localized function (ELF) of Hp-BN. (a) Atomic structure of Hp-BN with a super cell of 2*2*1. The dashed line indicates the unit cell. Red and blue balls present the B and

N atoms, respectively. (b), (c), (d) Slices of the ELF of Hp-BN in the [1 0 0], [1 -1 0], [0 1 0] direction, with the ZBNNR-like edges in the plane. The reference scale for the value of ELF is provided at the right bar.

The stability of the Hp-BN structure can be first understood by analyzing its deformation electron density. Our Mulliken population analysis[30] reveals an electron transfer of 0.56 *e* from each B to N atom, mainly in the *s* and *p* orbitals. Because the transferring electron of Hp-BN is 0.05 *e* smaller than that of c-BN and w-BN, the B-N bond length and cohesive energy of Hp-BN are much larger. To further identify the chemical bond characters, we calculated the electron localized function (ELF) of Hp-BN. Slices of the ELF parallel to the (1 0 0), (0 1 0) and (1 -1 0) crystal surfaces crossing the B1-N1, B3-N3 and B2-N2 planes are plotted in Figure 2b-2d, respectively. It is found that the ELF value on the N (B) atom is ~ 0.5 (~ 0.25), suggesting fully delocalized electrons (very low charge density), while the value between the B and N atoms is close to 0.9, indicating the formation of strong B-N σ bonds with fully localized electrons. Further analysis on the projected atomic DOS in Figure S1 of the Supporting Information shows that the *s* and *p* orbitals of B and N atom overlap in a broad range below the Fermi level, suggesting a strong bonding state with distorted $sp^3$ hybridization.

The dynamical stability of Hp-BN is examined by the phonon dispersion displayed in Figure 3 along several high-symmetry directions, together with the corresponding phonon DOS. There is no imaginary phonon frequency in the entire Brillouin zone, confirming that Hp-BN is dynamically stable. The highest optical branch of Hp-BN is located between Γ and H with a frequency of up to 1155 $cm^{-1}$. As there are three BN formula units in the primitive cell, the phonon dispersion of Hp-BN is much more complicated than that of c-BN and w-BN, which has only 6 and 12 phonon bands, respectively.[35, 36]

In the following, we will take the Brillouin zone center G as an example to study the phonon properties of Hp-BN. Aside from the three-fold degenerated acoustic branches, the other phonon bands of Hp-BN are divided into five two-fold degenerated bands and five non-degenerated ones at G. The highest optical branch of Hp-BN is two-fold degenerated at 1114 cm$^{-1}$ (close to c-BN and w-BN), corresponding to the out-of-plane vibration or stretching mode of B1-N1 and B2-N2, and to the bending mode of B3-N3 as shown in Figure 3d. Analysis on the partial phonon DOS reveals that in the high frequency region, the vibration frequency of Hp-BN mainly originates from B atoms because of their relative small mass. While in the frequency window from 700 cm$^{-1}$ to 850 cm$^{-1}$, the contributions from B and N are almost equivalent. For the vibration mode of a nearly flat band around 704 cm$^{-1}$, we found that it consists of three in-plane vibrations, i.e., B1-N2 along [1 0 0], B2-N3 along [0 1 0] and B3-N1 along [-1 -1 0].

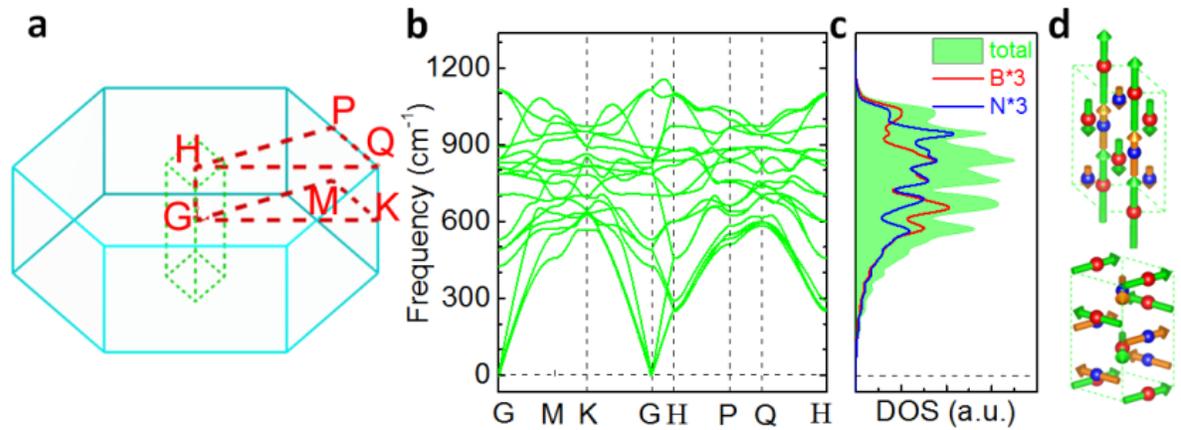

**Figure 3** Dynamical stability of Hp-BN. (a) High symmetry directions in the Brillouin zone. (b) Phonon band structure and (c) the total DOS associated (shadow). The inset shows the high-symmetry directions in the Brillouin zone. We have plotted the projected DOS of B and N atoms with red and blue lines, respectively. (d) Corresponding phonon vibration modes at 1114 cm$^{-1}$ (top) and 704 cm$^{-1}$ (bottom).

Many allotropes of BN structures have excellent mechanical properties.[37-39] For example, c-BN is the second hardest material known, inferior only to diamond. Since Hp-BN has a simple orthorhombic unit cell belonging to the P6$_2$22 space group, the 9 independent elastic constants $C_{11}$, $C_{22}$, $C_{33}$, $C_{44}$, $C_{55}$, $C_{66}$, $C_{12}$, $C_{13}$ and $C_{23}$ obtained at PBE level are listed in Table 1. To ensure the reliability of our DFPT results, we list the calculated elastic constants of c-BN and w-BN in the table as well. Our results for c-BN and w-BN are consistent with previous calculations, which suggest that the elastic constants obtained here are reliable. Moreover, the necessary and sufficient Born criteria for a 3D material with simple orthorhombic structure is $C_{11} > 0$, $C_{11}C_{22} > C_{12}^2$; $C_{11}C_{22}C_{33} + 2C_{12}C_{13}C_{23} - C_{11}C_{23}^2 - C_{22}C_{13}^2 - C_{33}C_{12}^2 > 0$; $C_{44} > 0$; $C_{55} > 0$; $C_{66} > 0$. [40] Clearly, the elastic constants of Hp-BN satisfy these mechanical stability criteria. It is also noted in Table 1 that the elastic constants $C_{ii}$ (i = 1, 2, 3) of Hp-BN are larger than those of c-BN, implying that Hp-BN has larger Young's modulus and higher stiffness. Meanwhile, the elastic constant $C_{33}$ is 11% larger than $C_{11}$ and $C_{22}$, suggesting that the out-of-plane B-N bonds are much stronger than the in-plane ones. By fitting the Birch-Murnaghan 3$^{rd}$-order EOS,[41] it turns out that the bulk modulus *B* of Hp-BN is 375 GPa, comparable to the values of c-BN (373 GPa) and w-BN (376 GPa), indicating that Hp-BN would be one of the hardest materials with excellent mechanical properties for engineering applications.

**Table 1** Comparison of space groups, lattice constants, elastic constants, bulk moduli and energy gaps of c-BN, w-BN and Hp-BN. ([a] Ref. 37; [b] Ref. 39.)

|  |  | c-BN | w-BN | Hp-BN |
|---|---|---|---|---|
| Space group | | No.216($T_d$) F43m | No.186($C_{6v}$) P6$_3$mc | No.180($D_6$) P6$_2$22 |
| Lattice (A) | | a = 3.624 | a = 2.554 c = 4.224 | a = 2.610 c = 5.828 |
| Elastic constant (GPa) | $C_{11}$ | 810; 820[a] | 960; 944[b] | 892 |
|  | $C_{12}$ | 181; 190[a] | 142; 149[b] | 166 |
|  | $C_{13}$ | - | 67; 83[b] | 109 |
| | $C_{22}$ | - | - | 890 |
| | $C_{23}$ | - | - | 110 |
| | $C_{33}$ | - | 1049; 1011[b] | 1015 |
| | $C_{44}$ | 454; 480[a] | 344; 347[b] | 363 |
| | $C_{55}$ | - | - | 341 |
| | $C_{66}$ | - | 409; 401[b] | 341 |
| Bulk Modulus (GPa) | | 373 | 376 | 375 |
| Energy gap (eV) | | 4.49 | 5.03 | 3.45 |

Next, we will focus on the electronic properties of Hp-BN. The band structure, calculated by standard DFT calculations within GGA at PBE level, is plotted in Figure 4. The band structure of Hp-BN along G-M-K-G is different from that along H-P-Q-H, exhibiting anisotropic electronic properties. The conduction band minimum (CBM) and valance band maximum (VBM) are located at the M and G points, respectively. The energy gap between CBM and VBM is 3.45 eV, which is much smaller than that of c-BN and w-BN as listed in Table 1. As DFT calculations always underestimate the energy gap, we performed a hybrid HSE06 functional calculation, in which an additional Hartree-Fock contribution is included in the short-range part of the generalized gradient approximation. A much wider indirect energy gap of 4.95 eV was obtained, and the VBM and CBM are shifted upward by 0.36 eV and 1.86 eV, respectively. To better understand the electronic properties of Hp-BN, we plotted the band decomposed DOS for the *s* and *p* orbitals in Figure 4b and 4c, respectively. It is found that, in the energy window from -12 eV to 10 eV, the main contribution from s orbitals lies about 6 eV (7 eV) above (below) the Fermi level, while in the low energy region, the electronic states are mainly composed of p orbitals. Hence, we conclude that the low energy electronic properties of Hp-BN are dominated by the p orbitals of B and N atoms.

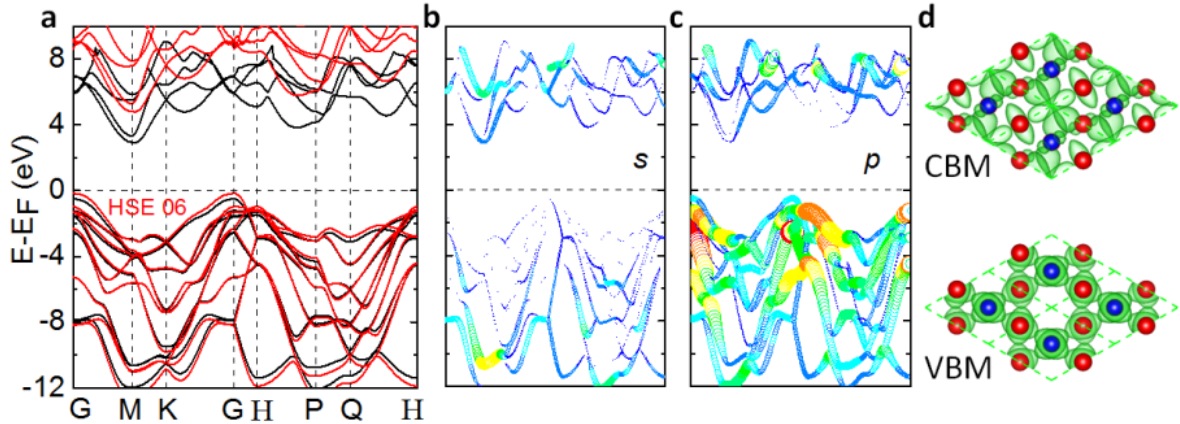

**Figure 4** Electronic properties of Hp-BN. (a) Band structure of Hp-BN along the high-symmetry *k* points in the Brillouin zone. Black and red lines represent the PBE and HSE06 results, respectively. (b) (c) Band decomposed DOS of Hp-BN with *s* and *p* characters plotted. (d) Top view of charge density distribution for CBM and VBM with the isosurface of 0.02 $e/Å^3$.

To further study the semiconducting properties of Hp-BN, we plotted the charge density distribution for the CBM and VBM in Figure 4d. Along the ***c*** axial, CBM exhibits rotated in-plane *p* orbital characters of B atoms while VBM exhibits distorted out-of-plane *p* orbital characters of N atoms. With the insight of the projected atomic DOS, we can conclude that the main contributions to the CBM and VBM are given by the distorted $p_x$ orbitals of the B atom and the distorted $p_z$ orbitals of the N atom. Moreover, the effective mass (EM) of electrons and holes at the CBM and VBM are calculated to be 0.54 $m_0$ and 1.15 $m_0$, indicating that the intrinsic carrier mobility could be similar as c-BN and w-BN.

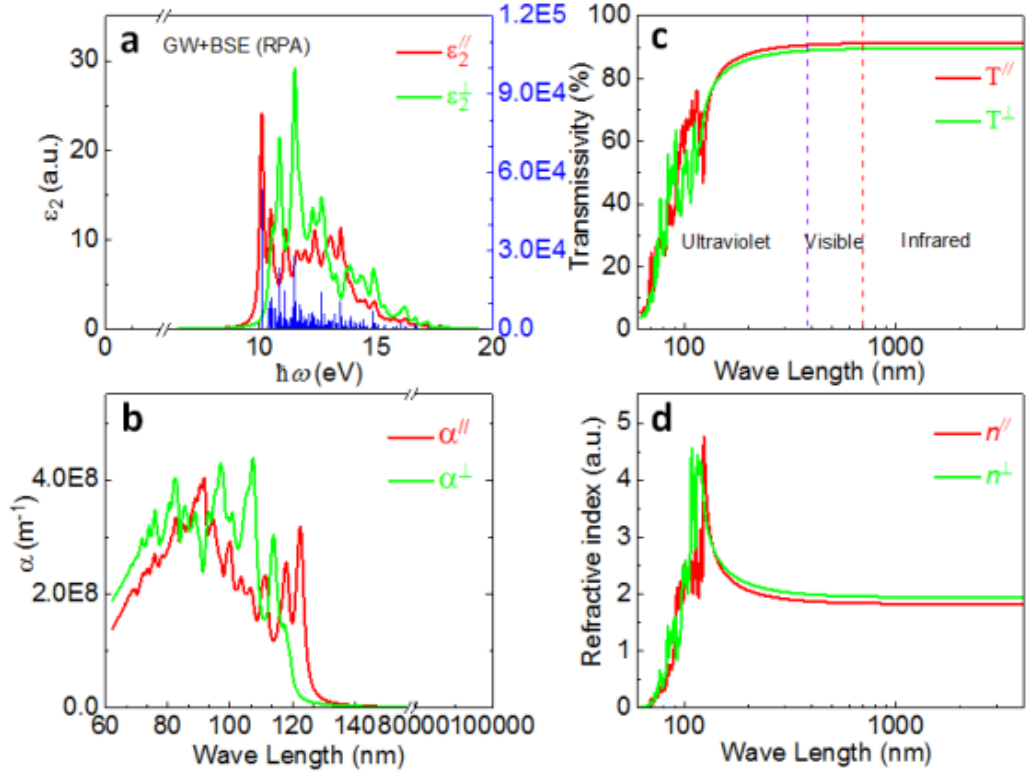

**Figure 5** Optical properties of Hp-BN. (a) Imaginary part of the dynamical dielectric function $\varepsilon_2(\omega)$ as a function of the photon energy $\hbar\omega$ for Hp-BN within random phase approximation (RPA). Vertical (blue) bars are the relative oscillator strengths for the optical transitions. The first lowest-energy peak at 10.08 eV in the spectrum corresponds to the first exciton of the in-plane oscillator. (b) Absorption coefficient, (c) transmission and (d) refractive index of Hp-BN. Red and green lines represent the transversal and vertical components of the optical coefficients, respectively.

We further studied the optical properties of Hp-BN within the GW approximation by solving the BSE of the two-particle Green's function. As optical absorption is dominated by the imaginary part of the frequency-dependent dielectric constant $\varepsilon(\omega) = \varepsilon_1(\omega) + i\varepsilon_2(\omega)$, we plotted the imaginary dielectric function $\varepsilon_2$ in the long wavelength limit $q \to 0$ in Figure 5a, as well as the optical transitions. Because of the symmetry of the crystal lattice, Hp-BN exhibits an anisotropic in-plane and out-of-plane imaginary dielectric function, where $\varepsilon_2^{//}$ and $\varepsilon_2^{\perp}$ represent the transversal and vertical part of the dielectric tensor, respectively. It is noted that both $\varepsilon_2^{//}$ and $\varepsilon_2^{\perp}$ show strong response to the photon energy from 9 eV to 15 eV. The first lowest-energy exciton of the optical transition function with the highest oscillator strength corresponds to the first peak of the transversal dielectric function $\varepsilon_2^{//}$ at 10.08 eV, indicating that the electrons of in-plane orbitals ($p_x/p_y$) are much easier to be excited comparing to the out-of-plane orbitals ($p_z$).

More measurable optical coefficients of Hp-BN, including the absorption coefficient α, refractive index and transmissivity are plotted in Figure 5b, 5c and 5d, respectively. Hp-BN shows broad absorption for photons with wavelength between 60-120 nm with a magnitude order of $10^8$ m$^{-1}$, suggesting its great potential in ultraviolet light absorption applications. On the other hand, Hp-BN has a high transmissivity of up to 91% for photons with wavelength larger than 0.15 μm, which includes the whole visible and infrared region, indicating that Hp-BN is colorless and transparent. Within the transparency range, the absorption coefficient is very small and the refractive index could have only a real component with no imaginary part. The calculated refractive index value of the birefringence is approximately equal to 1.84 for the transversal propagation of photons and 1.96 for vertical components. Considering the fact that Hp-BN is transparent for the visible and infrared light with huge ultraviolet absorption, it can be used in the optical windows and filters.

**Conclusion**

In conclusion, a comprehensive first-principles DFT study of 3D BN crystal with hyperhoneycomb structure is performed. We show that Hp-BN is a wide gap semiconductor with fully distorted $sp^3$-like hybridization. It is both dynamically and mechanically stable, and the calculated bulk modulus is even higher than that of c-BN. The carriers at CBM and VBM are separated at different Brillouin zone points, which prevent the recombination of electron and hole, indicating possible applications in energy storage devices. Moreover, the optical properties of Hp-BN are anisotropic, it has birefringence characters and shows broad absorption in the ultraviolet region. Meanwhile, it is transparent in the visible and infrared region, which makes it a possible candidate in many optical applications. Thus, we hope that the present theoretical prediction will inspire considerable experimental enthusiasm.

**Supporting information**

The Wyckoff positions and projected atomic density of states for Hp-BN.

**Author information**

Correspondence should be addressed to S. Y. (s.yuan@ science.ru.nl).

**Acknowledge**

Yu and Yuan acknowledge financial support from NSAF U1530401 and computational resources from the Beijing Computational Research Center. Katsnelson and Yuan acknowledge financial support from the European Research Council Advanced Grant program (Contract No. 338957).